\magnification=1200 \vsize=8.9truein \hsize=6.5truein \baselineskip=0.6truecm
\parindent=1truecm \nopagenumbers \font\scap=cmcsc10 \hfuzz=0.8truecm
\font\tenmsb=msbm10
\font\sevenmsb=msbm7
\font\fivemsb=msbm5
\newfam\msbfam
\textfont\msbfam=\tenmsb
\scriptfont\msbfam=\sevenmsb
\scriptscriptfont\msbfam=\fivemsb
\def\Bbb#1{{\fam\msbfam\relax#1}}

\def\xup{\overline x}
\def\xd{\underline x}

\def\uu{\overline{u}}
\def\uuu{\mathop{\overline{\overline{{u}}}}}

\def\wuu{\mathop{\overline{\overline{{w}}}}}
\def\adj{\rm adj}
\def\wu{\overline{w}}

\null \bigskip  \centerline{\bf Discrete and Continuous
Linearizable Equations}

\vskip 2truecm
\centerline{\scap S. Lafortune$^{\dag}$}
\centerline{\sl LPTM et GMPIB,  Universit\'e Paris VII}
\centerline{\sl Tour 24-14, 5$^e$\'etage}
\centerline{\sl 75251 Paris, France}
\footline{\sl $^{\dag}$ Permanent address: CRM, Universit\'e de
Montr\'eal, Montr\'eal, H3C 3J7 Canada}
\bigskip
\centerline{\scap B. Grammaticos}
\centerline{\sl GMPIB, Universit\'e Paris VII}
\centerline{\sl Tour 24-14, 5$^e$\'etage}
\centerline{\sl 75251 Paris, France}
\bigskip
\centerline{\scap A. Ramani}
\centerline{\sl CPT, Ecole Polytechnique}
\centerline{\sl CNRS, UMR 7644}
\centerline{\sl 91128 Palaiseau, France}
\bigskip\bigskip

Abstract
\smallskip \noindent
We study the projective systems in both continuous and discrete settings.
These systems
are linearizable by construction and thus, obviously, integrable. We show
that in the
continuous case it is possible to eliminate all variables but one and
reduce the system
to a single differential equation. This equation is of the form of those
singled-out by
Painlev\'e in his quest for integrable forms. In the discrete case, we
extend previous
results of ours showing that, again by elimination of variables, the
general projective
system can be written as a mapping for a single variable. We show that this
mapping is a
member of the family of multilinear systems (which is not integrable in
general). The
continuous limit of multilinear mappings is also discussed.
\vfill\eject

\footline={\hfill\folio} \pageno=2

\bigskip
\noindent {\scap 1. Introduction}
\medskip

The study of higher-order integrable systems is an interesting and open
problem. It is all the more
interesting when we realize that ``higher'' in this context means ``higher
then two''. As a matter
of fact the only instance where a complete classification of integrable
systems can be given is in
the case of second order differential equations. The study of equations of
the form:
$$
w''=f(w',w,z) \eqno(1.1)
$$
(where $f$ is polynomial in $w'$, rational in $w$ and analytic in
$z$) by
Painlev\'e [1] and Gambier[2], based on the singularity structure of the
solutions
of (1.1), led
to the complete classification
of integrable equations of this type. Integrable in this setting means
equations that
can either
\item{a)} be integrated through quadratures
\item{b)} be reduced to a linear differential system through some local
transformation
\item{c)} be integrated by isospectral methods involving linear
integrodifferential equations.

\noindent  Calogero has coined the names C- and S-integrability for the two
first types and
third type of integrability
respectively [3].

It is interesting to point out here that precisely the same
three types of integrability were
encountered when we undertook the the study of integrability of
three-point
non autonomous mappings
of the form:
$$
\xup={f_1(x)-f_2(x)\xd \over f_4(x)-f_3(x)\xd} \eqno(1.2)
$$
where $f_i$ are polynomial in $x$ [4]. The main guide was the study of the
singularity structure of the
solutions of (1.2). However, contrary to the continuous case, the complete
classification of the
integrable forms of (1.2) does not exist yet, although we are in possession
of (at least) one discrete
equation, for every member of the Painlev\'e/Gambier classification. The
situation is further complicated in the
discrete case by the fact that there exist two different kinds of discrete
equations. As a matter
of fact discrete  equations can be of either additive or multiplicative
type [5]. In the first case, the
independent discrete variable appears linearly while in the second one the
dependence is exponential.

In previous works [6,7,8] we have tried to  extend our results on integrable
second-order systems to
equations of higher order. A large class of equations which are amenable to
treatment is that
corresponding to the second kind of integrability introduced above, namely
integrability through
linearisation.
In particular we have concentrated on the discrete analogues of
linearisable equations. In [9] we have
studied the  mapping trilinear in the $x$'s:
$$
a_1\overline{x}_1x_1\underline{x_1}+a_2\overline{x}_1x_1+a_3\overline{x}_1
\underline{x_1}+a_4\overline{x}_1+
b_1 x_1\underline{x_1}+b_2x_1+b_3\underline{x_1}+b_4=0 \eqno(1.3)
$$
and have shown that it contains as a subcase a mapping linearisable by
reduction to a linear system.
This case turned out to be the $N$=2 discrete projective Riccati equation
that we introduced in [8], in
the general $N$-dimensional case. Other approaches to higher-order systems
can be found in [6].

In this paper we shall examine again the projective family of linearisable
systems. We shall show in
particular that it is possible to express the projective Riccati system as
a single equation  i.e. an
equation for a single dependent variable, both in the continuous and in the
discrete case. In the
latter case the form of the system is a multilinear generalisation of
(1.3). (A word of caution is due
here. The terms ``bi'', ``tri'' or ``multi''-linear should not be confused
with the Hirota
terminology. In the latter bilinear means a homogeneous quadratic
expression while bilinear in our
case means an expression where every variable enters linearly up to a
highest degree of homogeneity
two.) The multilinear mapping is, of course, much more general than the
mere linearisable projective
one. We illustrate this by studying some selected low-dimensional cases.
Moreover we indicate how one
can obtain the continuous limit of a multilinear mapping in a fairly
general setting.

\bigskip
\noindent {\scap 2. Continuous Projective Systems}
\medskip

We begin here by recalling what the continuous projective system is [10]. Our
starting point will be the
following  first order linear ODE for the
$(N+1)$-components vector $u$:
$$
u'=Cu \eqno(2.1)
$$
where $C$ is an $(N+1)\times (N+1)$ matrix depending on the independent
variable $z$. We define
the projective variables
$$
w_i=u_i/u_{N+1},\;\;i=1,\dots,N. \eqno(2.2)
$$
We then have $$w'_i={u'_i \over u_{N+1}}-{u_iu'_{N+1}\over u_{N+1}^2}$$
which can be written in
terms of the projective variables only. Thus the projective system reads:
$$
w'_i=(\sum_{j=1}^{N} C_{ij}w_j+C_{iN+1})-w_i(\sum_{j=1}^{N}
C_{N+1j}w_j+C_{N+1N+1}),\;\;1\leq
i\leq N. \eqno(2.3)
$$
For the $N=1$ case, we get the Riccati equation
$$
w'=aw^2+bw+c \eqno(2.4)
$$
where $a$, $b$ and $c$ are functions of the independent variable $z$. This
equation is
special as far as singularity analysis is concerned. If we consider a first
order equation of the form
$$
w'=f(w,z) \eqno(2.5)
$$
where $P$ is a polynomial in $w$ and analytic in $z$, the only nonlinear
equation of the form (2.5)
satisfying the Painlev\'e property is the Riccati.

In the case of the $N=2$ projective system, we can eliminate $w_2$ in the
system (2.3). The
equation then reads:
$$
(b_1
w_1'+b_2)w_1''+b_3w_1'^2+(b_4w_1+b_5)w_1'+b_6w_1^3+b_7w_1^2+b_8w_1+b_9=0
\eqno(2.6)
$$
where the $b_k$'s are functions of the $C_{ij}$'s.
This is an equation which corresponds to one of the fifty equations of the
Painlev\'e-Gambier classification. After a suitable change of variable, we
write it in its canonical form [11]:
$$
w''+3ww'+w^3+\phi(z)(w'+w^2)=0. \eqno(2.7)
$$
where $\phi$ is a free function of the independent variable $z$.
 We will now show that, in general, the continuous projective system can be
written as one single equation.  From (2.3), we rewrite the
first-order equation for $w_1$:
$$
w'_1=\sum_{j=1}^{N} C_{1j}w_j+C_{1N+1}-w_1(\sum_{j=1}^{N}
C_{N+1j}w_j+C_{N+1N+1}). \eqno(2.8)
$$
Then, differentiating both sides of (2.8),   and keeping track of the
$w_j$'s, $j\neq 1$
$$
w''_1=\sum_{j=2}^{N}
C_{1j}'w_j+\sum_{j=2}^{N}C_{1j}w_j'-w'_1(\sum_{j=1}^{N}
C_{N+1j}w_j)$$
$$-w_1(\sum_{j=2}^{N}
C_{N+1j}w_j'+\sum_{j=2}^{N}
C_{N+1j}'w_j)+D(w'_1,w_1,z)\eqno(2.9)
$$
where $D$ depends only on $w'_1$, $w_1$ and the independent variable $z$. We
see here that the only terms with degree higher than linear in the $w_k$'s,
$k\neq$1, are the $w_j'$ terms. But these appear only through the combination
$$
\sum_{j=2}^{N}( C_{1j}-w_1
C_{N+1j})w_j'\eqno(2.10)
$$
 Using (2.3) and keeping only the potentially nonlinear terms coming from
the second term in the lhs, we see that we should only worry about an
expression of the form: $\left(\sum_{j=2}^{N}(C_{1j}-w_1
C_{N+1j})w_j\right) \left(\sum_{k=1}^{N}
C_{N+1k}w_k+C_{N+1N+1}\right)$
which is, potentially, nonlinear in the $w_k$'s, $k\neq$1.
Note however,
that the first factor can be expressed in terms of $w'_1$ and $w_1$ only,
since this is precisely the combination of $w_j$'s that enter in (2.8).
Thus by replacing this factor by its expression in terms of $w'_1$ and
$w_1$ the only source of $w_k$'s, $k\neq 1$, comes from the second factor and
thus no nonlinearities are introduced.

Regrouping all terms we find:
$$w_1''=\sum_{j=1}^Nf_{2j}(w_1,w'_1,z)w_j+f_{2N+1}
(w_1,w'_1,z).
\eqno(2.11)$$
where the
$f_{2j}$'s are some definite functions of
$w_1$,
$w'_1$ and
$z$. In the general case, it is
not difficult to see that
$$
w_1^{(n)}=\sum_{j=1}^Nf_{nj}(w_1,w'_1,\dots,w_1^{(n-1)},z)w_j+f_{nN+1}
(w_1,w'_1,\dots,w_1^{(n-1)},z),\;\;n\in
\Bbb{Z}.
\eqno(2.12)
$$
So, we have a linear and inhomogeneous equation for the vector $W$ with
components
$w_1$, $w_2$,\dots,$w_N$:
$$
DW+E=0 \eqno(2.13)
$$
where $D$ is an $N\times N$ matrix and $E$ is an $N$ components vector
defined by
$$
\matrix{
D_{ij}=f_{ij} \cr\cr
E_i=f_{iN+1}-w_1^{(i)}.
}
\eqno(2.14)
$$
Hence, the $N^{\rm th}$ order equation satisfied by $w_1$ reads
$$
\sum_{j=1}^N {\adj{(D)}}_{1j}(w_1^{(j)}-f_{jN+1})-\det{(D)}w_1=0. \eqno(2.15)
$$
Thus the general projective system can indeed be written as an equation for a
single variable. For instance, in the case $N=3$, we get the following third
order equation [12]
$$
w'''={3/2\;{w''}^2+w''(g_1(z)ww'+g_2(z)w'+P_1(w))+P_3(w,w')+g_3(z)w^4
\over w'+P_2(w)}
\eqno(2.16)
$$
where $P_1(w)$ and $P_2(w)$ are polynomials of degree two in $w$,
$P_3(w,w')$ is a polynomial of degree three in $w$ and $w'$ and the $g_i$'s are
functions of the independent variable $z$. The polynomials have
coefficients depending on $z$.

For higher $N$'s, we can in principle obtain the corresponding $N$-th order
equation for a single
variable. However, the computations soon become unmanageable. Since we
cannot give a detailed general
expression, we must content ourselves with the global structure of this
equation. For order $N$, we have:
$$w^{(N)}=M(w^{(N-1)})^2+Pw^{(N-1)}+Q\eqno(2.17)   $$
where $M$, $P$ and $Q$ are rational functions of $w^{(N-2)}$,  $w^{(N-3)}$,
\dots, $w$, and analytic in $z$.  In particular, for $M$ we have:
$$M={1-1/n\over w^{(N-2)}+R} \eqno(2.18)$$
where $n$=1-$N$ and $R$ depends only on $w^{(N-3)}$, \dots, $w$, $z$.

What is interesting is that equation (2.17) together with (2.18) is exactly
one of the forms given by
Painlev\'e [13] for $N$-th order equations having the Painlev\'e property.
According to Painlev\'e,
when $M$ has just the simple expression (2.18) then $P$ and $Q$ must be, as
functions of
$w^{(N-2)}$, ratios of a polynomial of degree at most one for $P$ and at
most three for $Q$,
divided by the same denominator as that of $M$. We expect this property to
hold for our projective
systems, since they {\sl are} integrable and thus should have the
Painlev\'e property. We have
indeed checked that this holds true at least up to order 5.
\bigskip
\noindent {\scap 3. Discrete Projective Systems}
\medskip

In the discrete case, the projective system comes from the following
two-points linear system for
the $(N+1)$-component vector
$u$:
$$
\uu=Mu \eqno(3.1)
$$
where $M$ is an $(N+1)\times (N+1)$ matrix with elements depending on the
discrete independent variable $n$ and where we use the
notation
$u=u(n)$ and $\uu=u(n+1)$. We define the projective variables:
$$
w_i=u_i/u_{N+1},\;i=1,\ldots,N. \eqno(3.2)
$$
We then get the projective system [8]:
$$
\overline{w}_i={\sum_{j=1}^{N}M_{ij}w_j+M_{iN+1} \over
\sum_{j=1}^{N}M_{N+1j}w_j+M_{N+1N+1}}.\eqno(3.3)
$$
In the case $N=1$, we get the single equation
$$
\overline{w}_1={M_{11}w_1+M_{12} \over M_{21}w_1+M_{22}} \eqno(3.4)
$$
which is the discrete form of the Riccati equation. Mapping (3.4) is very
special as far as
integrability is concerned [9]. Consider mappings of the form
$$
\overline{v}=f(v,n) \eqno(3.5)
$$
where $f$ is some rational function in $v$. We can easily convince
ourselves that the only non-polynomial mapping of the
form (3.5) that has confined singularities is a mapping of the form
$$
\overline{v}=A_0+\sum_k{A_k \over (a_k v+d_k)^{m_k}}, \eqno(3.6)
$$
with $m_k$ integer. However if we consider the ``backward''
evolution towards diminishing $n$ then
(3.6),  $v$ expressed in terms of $\overline{v}$, is not rational. The only
nonlinear mapping
with confined singularities that is rational in both directions is just
$$
\overline{v}=A_0+{A_1 \over a_1 v+d_1}\equiv{av+b \over cv+d}. \eqno(3.7)
$$

In the $N=2$ case, we can write the projective system as a single
three-point mapping for $w_1$ [9]. To do this, we write
the equation (3.1) for $\uu$ and also the equation satisfied by
$\underline{u}$:
$$
\uu=Mu \eqno(3.8a)
$$
$$
\underline{u}=Pu \eqno(3.8b)
$$
where $M$ and $P$ are two $n$-dependent $3\times 3$ matrices and $P$
is given in terms of $M$ by
$$
P=\underline{M}^{-1}. \eqno(3.9)
$$
>From (3.8a) and (3.8b) we have the two equations
$$
\overline{w}_1={M_{11}w_1+M_{12}w_2+M_{13} \over
M_{31}w_1+M_{32}w_2+M_{33}} \eqno(3.10a)
$$
$$
\underline{w_1}={P_{11}w_1+P_{12}w_2+P_{13} \over
P_{31}w_1+P_{32}w_2+P_{33}}. \eqno(3.10b)
$$
>From the system (3.10) we easily write the three-point mapping satisfied
by $w_1$:
$$
a_1
\overline{w}_1w_1\underline{w_1}+a_2\overline{w}_1w_1+a_3\overline{w}_1
\underline{w_1}+a_4\overline{w}_1+
b_1 w_1\underline{w_1}+b_2w_1+b_3\underline{w_1}+b_4=0 \eqno(3.11)
$$
where the $a_i$'s and the $b_i$'s are functions of the components of $M$
and $P$. We note that (3.11) is  of the
form of the most general multilinear equation in $\overline{w}_1$, $w_1$
and $\underline{w_1}$ one can write.

In the general case, we will prove here that one can still write the
projective system as one single equation. We will
also see that the mapping we obtain has also the form of the most general
multilinear equation one can write in terms of
$w_1$ and its $N$ first upshifts.

For arbitrary $N$ we first write a system of the form (3.1) for $\uu$ and
the systems satisfied
by the $N$ first upshifts of $u$:
$$
\matrix{
\uu=M_1u \cr
\overline{\uu}=M_2u \cr
\vdots\cr
\displaystyle{\overline{\uuu^{:}}}=M_Nu
}\eqno(3.12)
$$
where the last object is the $N$-th upshift of $u$ and the $M_i$'s are
$(N+1)\times(N+1)$
$n$-dependent matrices. For
$i>2$, $M_i$ can be recursively
expressed in terms of $M_1$ and its upshifts through
$M_{i}=\overline{M}_{i-1}M_1$. We then get the following mappings for
$w_1=u_1/u_{N+1}$:
$$
\matrix{
\displaystyle{\overline{w}_1={\sum_{j=1}^{N}(M_1)_{1j}w_j+(M_1)_{1N+1}
\over \sum_{j=1}^{N}(M_1)_{N+1j}w_j+(M_1)_{N+1N+1}}}
\cr\cr
\displaystyle{\overline{\overline{w}}_1={\sum_{j=1}^{N}(M_2)_{1j}w_j+(M_2)_{1N+1
}
\over \sum_{j=1}^{N}(M_2)_{N+1j}w_j+(M_2)_{N+1N+1}}}
\cr\cr
\vdots
\cr\cr
\displaystyle{\overline{\wuu^:}_1={\sum_{j=1}^{N} (M_N)_{1j}w_j+(M_N)_{1N+1}
\over
\sum_{j=1}^{N}(M_N)_{N+1j}w_j+(M_N)_{N+1N+1}}} .}
\eqno(3.13)
$$
To write this system in a compact way, we define:
$$
\matrix{
v_{0}\equiv w_1 \cr\cr
v_{1}\equiv \overline{w}_1 \cr\cr
v_{2}\equiv \overline{\overline{w}}_1 \cr\cr
\vdots \cr
\displaystyle{v_{N}\equiv \overline{\wuu^{:}}}_1
}\eqno(3.14)
$$
and system (3.13) can be rewritten as
$$
v_{i}={\sum_{j=1}^{N}(M_i)_{1j}w_j+(M_i)_{1N+1} \over
\sum_{j=1}^{N}(M_i)_{N+1j}w_j+(M_i)_{N+1N+1}}\;
i=1,\dots,N.
\eqno(3.15)
$$
The system (3.15) is an inhomogeneous linear equation for the vector $W$
with components $w_j$, $j=1,\dots, N$:
$$
CW+B=0 \eqno(3.16)
$$
where $C$ is an $N\times N$ matrix and $B$ is an $N$-component vector
defined by
$$
\matrix{
C_{ij}=(M_i)_{1j}-v_{i}(M_i)_{N+1j} \cr\cr
B_i=(M_i)_{1N+1}-v_{i}(M_i)_{N+1N+1}
}\eqno(3.17)
$$
$$
1\leq i,j\leq N.
$$
Solving (3.16) for $w_1\equiv v_0$, we obtain:
$$
{\det{(C)}}v_0+\sum_{j=1}^{N}({\adj{(C)}})_{1j}B_j=0. \eqno(3.18)
$$
>From the definitions (3.17), we see that ${\det{(C)}}$ will
generically be
a general multilinear expression in the $v_i$'s (which are just the different
shifts of $w_1$). The first term of equation (3.18)  will
include all
the multilinear terms containing $v_0\equiv w_1$. Moreover
${\adj{(C)}}_{1j}$ is a general multilinear expression in
$v_i$ for $i=1,2,\dots,j-1,j+1,\dots,N$. The second term of
equation (3.18)
will thus be in the form of a general multilinear
expression in the shifts of $w_1$, the unshifted $w_1$ excepted. Hence, we
can see that
(3.18) has indeed the
form of the most general multilinear equation in
$w_1$ and its
$N$ first upshifts, except that the coefficients are not all free.

\bigskip
\noindent {\scap 4. Confinement of the projective Mapping}
\medskip

In this section we are considering the same projective system as
in Section 3.
 For the cases $N=2,3$ it is already known
that mapping (3.18) confines in one step [9].
The case $N=2$ simply corresponds to the Riccati mapping
(3.4). We will prove here that confinement [14] occurs in one step
for arbitrary
value of $N$.

First, we prove the following formula
$$
{ \partial( \wu_1,\wu_2, \dots, \wu_N) \over \partial(w_1, w_2,
\dots, w_N)}
=\det(M_1)\left ({u_{N+1}\over \uu_{N+1}}\right)^{N+1} \eqno(4.1)
$$
To prove this we first have to use the chain rule:
$$
\eqalignno{
{ \partial( \uu_1,\uu_2, \dots, \uu_{N+1}) \over
\partial(u_1, u_2, \dots, u_{N+1})}= \hskip8cm&\cr
{ \partial( \uu_1,\uu_2, \dots, \uu_{N+1}) \over
\partial( \wu_1,\wu_2, \dots, \wu_N,\uu_{N+1})}
{\partial( \wu_1,\wu_2, \dots, \wu_N,\uu_{N+1})
\over \partial(w_1, w_2, \dots, w_N,u_{N+1})}
{\partial(w_1, w_2, \dots, w_N,u_{N+1}) \over
\partial(u_1, u_2, \dots, u_{N+1})}
&\quad\quad\quad\quad\quad(4.2)}
$$
Then using (3.2) and (3.3), we easily calculate that
$$
{ \partial( \uu_1,\uu_2, \dots, \uu_{N+1}) \over
\partial( \wu_1,\wu_2, \dots, \wu_N,\uu_{N+1})}=\uu_{N+1}^N
\eqno(4.3)
$$
$$
{\partial( \wu_1,\wu_2, \dots, \wu_N,\uu_{N+1}) \over
\partial(w_1, w_2, \dots, w_N,u_{N+1})}=
{\partial( \wu_1,\wu_2, \dots, \wu_N) \over
\partial(w_1, w_2, \dots, w_N)}{\uu_{N+1}\over u_{N+1}}
\eqno(4.4)
$$
$$
{\partial( w_1,w_2, \dots, w_N,u_{N+1}) \over
\partial(u_1, u_2, \dots, u_N,u_{N+1})}
=u_{N+1}^{-N}
\eqno(4.5)
$$
Equation (4.1) is then proven.
But $M_1$ can never be singular. Indeed, if it were
singular for some $n$, then the mapping would always lose a degree
of freedom at $n$, independently of the initial conditions. This
would be a `fixed' singularity, which we do not consider. Hence,
in what follows, we will always consider that the Jacobian
between the
$\wu_i$'s and the
$w_i$'s is nonzero: the mapping giving the $\wu_i$'s from the
$w_i$'s is never singular.

What we mean by a singularity in the $v$-mapping is that the
information in the variables of the mapping is not the full one,
i.e. there is a loss of information at step $N$.

So  saying that the singularity enters with the set $(v_1,v_2, \dots,
v_{N})$, (which is the first one where $v_N$ appears) means
$$
{ \partial( v_1, v_2, \dots, v_{N}) \over
\partial(v_0(\equiv w_1),v_1, v_2, \dots, v_{N-1})} = (-1)^N
{\partial v_N\over\partial v_0}= 0
\eqno(4.6)
$$
Using the chain rule as well as (4.1) this writes:
$$
{ \partial( v_1, v_2, \dots, v_{N}) \over
\partial(\wu_1,\wu_2,  \dots, \wu_N)} \det M_1 \left ({u_{N+1}\over
\uu_{N+1}}\right)^{N+1}\left({\partial(v_0,v_1, v_2, \dots,
v_{N-1})\over\partial(w_1,w_2,  \dots, w_N)}\right)^{-1}=0
\eqno(4.7)
$$
Thus a singularity at step $N$ implies that:
$$
{\partial( v_1, v_2, \dots, v_{N}) \over
\partial(\wu_1,\wu_2,  \dots, \wu_N)}=0 \eqno(4.8)
$$
(note that $v_1=\wu_1$). We assume that the singularity actually
 occurs there for the first time i.e. that:
$$
{\partial(v_0, v_1, v_2, \dots, v_{N-1}) \over
\partial(w_1,w_2,  \dots, w_N)}\neq0 \eqno(4.9)
$$
Thus the appearence of the singularity is related to the fact that
the $N$ quantities $( v_1, v_2, \dots, v_{N})$ do not carry enough
information on the $(\wu_1,\wu_2,  \dots, \wu_N)$. Though the
$w_i$'s have all the information at all steps, some is lost in the
particular combination $( v_1, v_2, \dots, v_{N})$.
At step $N+1$ the relevant Jacobian does not vanish.
$$
{\partial( v_2, \dots, v_{N},v_{N+1}) \over
\partial(\overline{\wu}_1,\overline{\wu}_2,  \dots, \overline{\wu}_N)}\neq
0 \eqno(4.10)
$$
Indeed, this Jacobian contains information about $M_{N+1}$ and thus the
$N$-th upshift
of $M_1$, which is independent of all others, through, and only through
$v_{N+1}$. Thus since this quantity is independent from the others, the
Jacobian cannot
vanish. Formally, one should write the upshift of (3.18) as a multilinear
equation of
order $N+1$ in the $v_{1}$, $v_{2}$, \dots, $v_{N+1}$ and solve for
$v_{N+1}$ as:
$$v_{N+1}= {F(v_{1},v_{2},\dots, v_{N})\over G(v_{1},v_{2},\dots,
v_{N})}\eqno(4.11)$$
where $F$ and $G$ are multilinear functions of their arguments. But if
$v_{N+1}$ were
actually a function of $(v_{1},v_{2},\dots,v_{N})$, then it would follow
that the
$(v_{1},v_{2},\dots, v_{N+1})$ would not contain more information than
$(v_{1},v_{2},\dots, v_{N})$ and thus ($v_{2},\dots, v_{N+1})$ certainly
not more either,
in contradiction with (4.10). It follows that if (4.8)  holds, (4.11) must
be of the
form 0/0 (i.e. the upshift of (3.18) is identically zero as a function of
$v_{N+1}$).
In fact, (4.10) implies that the lost degree of freedom has been recovered
and that the
singularity is confined already at this step.

In the case $N=2$, the downshift of (4.8) reduces to
$$
v_0((M_1)_{11}(M_1)_{32}-(M_1)_{12}(M_1)_{31})+
(M_1)_{13}(M_1)_{32}-(M_1)_{12}(M_1)_{33}=0
\eqno(4.12)
$$
This implies that
$$
v_1=\overline w_1={(M_1)_{12} \over (M_1)_{32}}
\eqno(4.13)
$$
and thus does not depend on $\underline w_1$.
In the case $N=2$, (3.15) is only a system of two equations. The second
equation of (3.15) gives $v_2$ in terms of $v_1$ and $v_0$:
$$\displaylines{
v_2\Big(v_1v_0\big( (M_1)_{32}(M_2)_{31}- (M_1)_{31}(M_2)_{32}\big) +
v_1\big((M_1)_{32}(M_2)_{33} - (M_1)_{33}(M_2)_{32}\big)\hfill\cr\hfill
 + v_0\big((M_1)_{11}(M_2)_{32} - (M_1)_{12}(M_2)_{31}\big)
+(M_1)_{13}(M_2)_{32}- (M_1)_{12}(M_2)_{33}\Big)\hfill\cr\hfill
 + v_1v_0\big((M_1)_{31}(M_2)_{12} - (M_1)_{32}(M_2)_{11}\big)
 + v_1\big((M_1)_{33}(M_2)_{12}- (M_1)_{32}(M_2)_{13}\big)\hfill\cr\hfill
 + v_0\big( (M_1)_{12}(M_2)_{11}- (M_1)_{11}(M_2)_{12}\big) + (
M_1)_{12}(M_2)_{13} -(M_1)_{13}(M_2)_{12}=0\quad(4.14)\cr}$$
Implementing (4.12) and (4.13) into (4.14), we find that the
coefficient of $v_2$ and the rest of the equation are
both zero. So, formally, $v_2$ is indeed $0/0$ and a detailed calculation
shows that it does
depend on $\underline w_1$.

\bigskip
\noindent {\scap 5.  Continuous limits of multilinear mappings}
\medskip
In the previous sections, we have seen that the projective systems  can always
be written as multilinear mappings of a single variable. Since the continuous
system can also be  written as a single differential equation, it is
interesting to investigate the continuous limit of the multilinear mappings. We
shall do this in a general setting i.e. without imposing to the mapping the
integrability constraints associated to its projective character.

Our main assumption is that the discrete variable $u$
coincides with the continuous variable $w$ at the continuous limit. This will
indeed turn out to be the right choice. We start by considering that
we have a mapping of order $N$ (i.e. a $(N$+1)-point mapping). At the
continuous limit, this would lead to an $N$-th order differential
equation. Moreover, the highest nonlinearity is of degree $(N+1)$. Given its
structure, the multilinear mapping is invariant under the transformation $u\to
1/u$.  It turns out that the `richest' terms in the mapping are the `middle'
ones: they have the largest number of terms of the same homogeneity. More
precisely,  if the order $N$ of the mapping is odd, the degree of the middle
term is ($N$+1)/2. If the order is even then there are two `middle' terms with
degree $N$/2 and $N$/2+1. It suffices in this case to consider one of them,
say the one of degree $N$/2. Indeed it turns out that the contribution of
the other `middle' term can be absorbed into that of the first term through a
translation and subsequent inversion of the continuous variable. Since these
`middle' terms have the highest nonlinearity combined with the highest degree
they are the ones that play the
dominant role in the differential equation obtained.

Let us make these considerations more precise by analyzing specific examples.

We start with the simplest possible case of order $N$=1, the discrete Riccati
equation. We have:
$$ \alpha \uu u+\beta \uu +\gamma  u +\delta =0\eqno(5.1)$$
Clearly the two `middle' terms $\{\uu$, $u\}$ can be combined to produce a
first derivative $w'$ and the nonlinearity comes from the first term $\uu u\to
w^2$.  Let us now consider the somewhat more interesting
cubic case corresponding to $N=2$
$$ \alpha \uu u\underline u+\beta \uu u+\gamma  \uu \underline u+\delta u
\underline u
+\zeta \uu+\theta u +\kappa \underline u +\lambda=0\eqno(5.2)$$
The continuous limit is obtained through $u\to w$, $\uu= w+\varepsilon w'+
\varepsilon^2 w''/2$,
$\underline u= w-\varepsilon w'+\varepsilon^2 w''/2$. We assume that at
leading order the
coefficients behave like: $\beta=-B/\varepsilon^2+{\cal O}(1/\varepsilon)$,
$\gamma=2B/\varepsilon^2+{\cal O}(1/\varepsilon)$,
$\delta=-B/\varepsilon^2+{\cal
O}(1/\varepsilon)$, $\zeta=A/\varepsilon^2+{\cal O}(1/\varepsilon)$,
$\theta=-2A/\varepsilon^2+{\cal O}(1/\varepsilon)$,
$\kappa=A/\varepsilon^2+{\cal
O}(1/\varepsilon)$. We find then that the second-degree terms lead to
$B(ww''-2w'^2)$ at
lowest order in $\varepsilon$, while the first-degree one give simply
$Aw''$. Translating
$w$ by $A/B$ we can absorb the $Aw''$ term into $B(ww''-2w'^2)$. We define now
$x=(w+A/B)^{-1}$ and obtain as only dominant term $Bx''$. The detailed
continuous limit of
(5.2) must take into account the precise form of the $\alpha, \beta,\dots $
coefficients.
It was performed in [9] where we have shown that the continuous limit is
$$x''+3xx'+x^3+q(z)(x'+x^2)+r(z)x+s(z)=0.\eqno(5.3)$$
This is precisely the form of the linearisable differential equation
obtained by Painlev\'e
at order two.

In the case of the $N=3$ multilinear mapping there exists only one
contribution at dominant
order. Its continuous limit, obtained along lines similar to those of the
$N=2$ case,
leads to a dominant term $w'''w'-{3\over 2}w''^2$. Since this is the only
dominant
contribution it is interesting to study the subdominant ones. We find three
groups of
terms: $aw'''(w^2-6ww'w''+6w'^3)+bw'''(w-3w'w'')+cw'''$. Again we can use
the translation
freedom and absorb the term in $c$ into the term in $a$. Then, inverting
$x=1/w$, the $a$
term disappears, regenerating a term in $c$. Thus, at dominant order, and
since the leading
term is invariant under inversion, we have for the $N=3$ multilinear mapping
the continuous limit $x'''x'-{3\over 2}x''^2+bx'''(x-3x'x'')+cx'''$.

The case of the $N=4$ and $N=5$ mappings can be treated along similar
lines. In the case of
$N=4$ we find that the dominant terms are
$w''''(w''w-2w'^2)-{4\over 3}w'''^2w+8x'''x''x'-6w''^3$ and
$w''''w''-{4\over 3}w'''^2$.
Again a translation of $w$ allows to absorb the second term into the first
and an inversion
of the (translated) $w$ leads to the simple form for the dominant term
$x''''x''-{4\over
3}x'''^2$. For $N=5$ there exists only one contribution at dominant order
which reads
$x^{(\rm v)}(x'''x'-{3\over 2}x''^2)-{5\over
4}x''''^2x'+5x''''x'''x''-{10\over 3}x'''^3$.
As expected this expression is invariant under inversion.

What is interesting in the continuous limits we examined above is that in
each case the
leading behaviour is the one predicted by Painlev\'e and which we
encountered in section 2.
This is not in disagreement with the fact that the multilinear mappings are
not integrable
in general. The conditions that led Painlev\'e to the form (2.17) are just
the first
conditions for the Painlev\'e property to hold and do not suffice for
integrability.
However we expect the precise continuous limits of the projective systems
to fall exactly
within the class of integrable equations of Painlev\'e type (which for
orders higher than
two have not been fully classified yet).

\bigskip
\noindent {\scap 6.  Conclusion}
\medskip
In this paper we have examined the alternate forms that can assume
projective systems of
general orders. We have shown that in both the continuous and the discrete
case, it is
possible to write a single $N$-th order equation for one of the dependent
variables. In the
continuous case the differential equation obtained satisfies the necessary
integrability
conditions established by Painlev\'e. This is, of course, no surprise,
since the projective
systems are linearizable and thus integrable by construction. In the
discrete case, the
mapping obtained belongs to the class of multilinear mappings, which goes
beyond the
linearizable systems. (The  multilinear mappings are not integrable in
general, but do
contain integrable subclasses). We have explicitly shown that the multilinear
projective mapping obtained does satisfy the singularity confinement
integrability
criterion. Finally, we have indicated the procedure for obtaining the
continuous limit of
these mappings and we have shown that the resulting forms fall within the
class obtained by
Painlev\'e.
 \bigskip
\noindent {\scap Acknowledgements}.
\smallskip
\noindent
The financial help of
the CEFIPRA, through the contract 1201-1, is gratefully acknowledged. S.
Lafortune
acknowledges three scholarships. For his Ph.D.: one from NSERC (National
Science and
Engineering Research Council of Canada) and one from FCAR (Fonds pour la
Formation des
Chercheurs et  l'Aide \`a la Recherche du Qu\'ebec). For his stay in Paris:
a  scholarship
from  ``Programme de Soutien de Cotutelle de Th\`ese de doctorat du
Gouvernement du
Qu\'ebec''.
\bigskip
{\scap References}
\smallskip
\item{[1]} P. Painlev\'e, Acta Math. 25 (1902) 1.
\item{[2]} B. Gambier, Acta Math. 33 (1910) 1.
\item{[3]} F. Calogero, in {\sl What is Integrability?}, ed. V. Zakharov,
Springer, New York,
(1990) 1.
\item{[4]} B. Grammaticos, A. Ramani, Meth. Appl. Anal. 4 (1997) 196.
\item{[5]} A. Ramani, B. Grammaticos and J. Hietarinta, Phys. Rev. Lett. 67
(1991) 1829.
\item{[6]} S. Lafortune,  B. Grammaticos, A. Ramani, Inv. Prob. 14 (1998) 287.
\item{[7]} B. Grammaticos, A. Ramani, {\sl  Continuous and discrete
linearisable systems: the
Riccati saga,} Lecture at the Montr\'eal 97 symposium.
\item{[8]} B. Grammaticos, A. Ramani and P. Winternitz, Phys. Lett. A 245
(1998) 382.
\item{[9]} A. Ramani, B. Grammaticos and G. Karra, Physica A 181 (1992) 115.
\item{[10]} R.L. Anderson, J. Harnad and P. Winternitz, Physica D4 (1982)
164-182.
\item{[11]} E.L. Ince, {\sl Ordinary differential equations}, Dover, New
York, 1956.
\item{[12]} J. Chazy, Acta Math.  34 (1911) 317.
\item{[13]} P. Painlev\'e, C. R. Acad. Sci. (Paris) 130 (1900) 1112.
\item{[14]} B. Grammaticos, A. Ramani and V. Papageorgiou, Phys. Rev. Lett.
67 (1991)
1825.
 \end